\documentclass[11pt]{article}

\usepackage{amsmath,amssymb,graphicx}
\usepackage{hyperref}

\usepackage{multicol,color,longtable}

\definecolor{darkred}{rgb}{0.65,0.15,0}
\hypersetup{pdfborder={0 0 0},linktocpage=true,colorlinks=true,urlcolor=darkred,citecolor=blue,linkcolor=darkred}

\definecolor{AV}{rgb}{0.65,0.0,0}
\definecolor{AK}{rgb}{0,0,1}
\definecolor{DK}{rgb}{0.6,0.4,0}

\newcommand{\nn}{\nonumber}

\newcommand{\reals}{\mathbb{R}}
\newcommand{\cx}{\mathbb{C}}

\newcommand{\lag}{\mathfrak{g}}
\newcommand{\lak}{\mathfrak{k}}
\newcommand{\lap}{\mathfrak{p}}

\newcommand{\lb}{\left[}
\newcommand{\rb}{\right]}

\newcommand{\cV}{\mathcal{V}}

\newcommand{\cZ}{\mathcal{Z}}
\newcommand{\cM}{\mathcal{M}}

\newcommand{\id}{1\!\!1}

\newcommand{\cf}{f}

\newcommand\cA{\mathcal{A}}

\newcommand\MM{\textrm{MM}}
\newcommand\EE{\textrm{E}}

\makeatletter

\@addtoreset{equation}{section}
\makeatother

\begin{document}

\thispagestyle{empty}

{\flushright {AEI-2012-193}\\[15mm]}

\begin{center}
{\LARGE \bf Inverse Scattering and\\[3mm] the Geroch Group}\\[10mm]

\vspace{8mm}
\normalsize
{\large  Despoina Katsimpouri${}^{1}$, Axel Kleinschmidt${}^{1,2}$\\[2mm] and Amitabh Virmani${}^1$}

\vspace{10mm}
${}^1${\it Max-Planck-Institut f\"{u}r Gravitationsphysik (Albert-Einstein-Institut)\\
Am M\"{u}hlenberg 1, DE-14476 Potsdam, Germany}
\vskip 1 em
${}^2${\it International Solvay Institutes\\
ULB-Campus Plaine CP231, BE-1050 Brussels, Belgium}

\vspace{15mm}

\hrule

\vspace{10mm}

\begin{tabular}{p{12cm}}
{\footnotesize
We study the integrability of gravity-matter systems in $D=2$ spatial dimensions with matter related to a symmetric space $\mathrm{G}/\mathrm{K}$ using the well-known linear systems of Belinski--Zakharov (BZ) and Breitenlohner--Maison (BM). The linear system of BM makes the group structure of the Geroch group manifest and we analyse the relation of this group structure to the
inverse scattering method of the BZ approach in general. Concrete solution generating methods are exhibited in the BM approach
in the so-called soliton transformation sector where the analysis becomes purely algebraic. As a novel example we construct the Kerr--NUT solution by solving the appropriate purely algebraic Riemann--Hilbert problem in the BM approach.}
\end{tabular}
\vspace{5mm}
\hrule
\end{center}

\newpage
\setcounter{page}{1}

\tableofcontents

\section{Introduction}

In late 1970s and early 1980s a large variety of solution generating techniques for the four-dimensional vacuum Einstein equations and Einstein-Maxwell equations were explored, for an overview see~\cite{Stephani:2003tm}. It was later realized
in the case of two commuting and hypersurface orthogonal Killing vectors that all these approaches are nothing but different manifestations of the integrability of the corresponding effectively two-dimensional system that can be exhibited by means of a linear system or Lax pair~\cite{BZ, Breitenlohner:1986um,BV}. Several authors made efforts to find interrelations between these different methods. Cosgrove~\cite{Cosgrove:1980fx} took a computational approach whereas Breitenlohner and Maison~\cite{Breitenlohner:1986um} concentrated on unraveling the group theoretical structure behind these techniques, taking up ideas of Geroch~\cite{Geroch1970} and Julia~\cite{Julia1981}. Other relevant work includes~\cite{KC,Julia:1982fs,Devchand,Nicolai:1987kz,Schwarz:1995af,Bernard:1997et,Lu:2007jc}.  In the present work, we revisit these studies to further elucidate the interrelation between the various techniques.

The best developed technique for generating solutions is that of Belinski and Zakharov~\cite{BZ,BV},  henceforth BZ for short and often called the inverse scattering method. It has been very successful in constructing novel solutions in both four- and five-dimensional vacuum gravity. The method involves some rather special adjustments to certain quantities before new physical solutions can be obtained.  However, there are problems when applying this method to different gravity-matter systems like those of interest in supergravity where the implementation of the same adjustments fails and it is not guaranteed that an inverse scattering transformation preserves all features required of a solution to the gravity-matter equations~\cite{Figueras:2009mc}. In the group theoretical framework of Breitenlohner and Maison (BM)~\cite{Breitenlohner:1986um} this problem does not arise since the solution generating transformations form the so-called Geroch group (an affine group) and by the group property any transformation will generate a new solution. The drawback of the BM method is that it is not as easy to implement and does not always operate directly on the physical quantities. Despite these shortcomings, the promise of the BM approach is that it can be applied to various general settings of interest.

In order to illustrate this point, we consider for concreteness $D=4$ gravity with a space-like and a time-like Killing vector such that the system is effectively two-dimensional and we are in the realm of so-called stationary axisymmetric solutions. The infinite-dimensional affine symmetry group can be viewed as the closure of two finite-dimensional symmetry groups that act on the space of solutions~\cite{Geroch1970,Breitenlohner:1986um}. The first one is typically called the Matzner--Misner group SL$(2,\reals)_\MM$ and consists of area preserving diffeomorphisms of the orbit space of the two Killing vectors (the `torus' that one reduced on, the Killing vector orbits are not necessarily compact). The other group is called the Ehlers group SL$(2,\reals)_\EE$ and is a hidden symmetry already of the three-dimensional model. The fields that it acts on are partly formed from dualising a Kaluza--Klein vector field (in $D=3$) to a scalar field; therefore it does not act directly on the metric components. Combining SL$(2,\reals)_\MM$ and SL$(2,\reals)_\EE$ yields the infinite-dimensional affine Geroch group~\cite{Geroch1970,Breitenlohner:1986um}.

The linear systems of BZ and BM display the integrability of the $D=2$ theory in
terms of generating functions that depend on spectral parameters besides the dependence on the two-dimensional space(-time) coordinates $x^m$. There are in principle four choices for the linear system and generating functions at this point that can be summarized in the following diagram:
\begin{center}
\begin{tabular}{r|ccc}
&Ehlers && Matzner--Misner\\[2mm]\hline\\
BZ & $\Psi_\EE(\lambda,x)$ &$\longleftrightarrow$& $\Psi_\MM(\lambda,x)$\\[3mm]
&$\updownarrow$&&$\updownarrow$\\[3mm]
BM & $\cV_\EE(t,x)$  &$\longleftrightarrow$& $\cV_\MM(t,x)$
\end{tabular}
\end{center}
The horizontal arrows in this diagram correspond to duality transformations between some of the two-dimensional scalar fields. At the level of the Lie group this corresponds to the action of an (outer) automorphism, now known as the Kramer--Neugebauer mapping~\cite{KN,Breitenlohner:1986um,Sen:1995qk}. The vertical lines relate the two linear systems and are the main subject of the present work. Note that the spectral parameters of the generating functions are different in the two rows. The traditional inverse scattering method of BZ operates on the physical variables and so is located in the Matzner--Misner column of the diagram. In fact it uses a slight extension of the Matzner--Misner variables in that it also treats the volume of the two-dimensional orbit space and so corresponds to GL$(2,\reals)_\MM$ rather than SL$(2,\reals)_\MM$. In our work, we will focus mainly on the left column where the group theoretical structures are a bit nicer, e.g., empty Minkowski space corresponds to the identity function. We will always display the subscripts $\EE$ and $\MM$ to avoid any confusion.

It follows from the work of BM that any gravity-coupled matter system in $D=3$ with symmetric space target G$_\EE$/K$_\EE$ becomes integrable when an additional Killing symmetry is present. Here, G$_\EE$ is the generalization of the Ehlers symmetry $\mathrm{SL}(2,\reals)_\EE$ to other systems. The group of symmetry transformations form the (non-twisted) affine extension of G$_\EE$ and we will restrict to simple G$_\EE$ for simplicity. Beyond this general fact, it is quite hard to construct actual solutions explicitly using this method since one has to solve a matrix valued Riemann--Hilbert problem~\cite{Stephani:2003tm, Breitenlohner:1986um, EH}, sometimes also referred to as Birkhoff factorization. In the case analogous to the soliton transformations of BZ, however, the problem reduces to a linear algebra system that can be solved as shown in~\cite{BMunpub} which we will review below since the work is not published.

Our chief motivation to study the different formulations of integrability is to find their interrelation and, by this, to make new methods available for generating solutions of (super-)gravity beyond the cases that have been covered so far. Of particular interest are cases when $\mathrm{G}_\EE$ is an exceptional symmetry group and other cases that arise in string theory. There are no known established systematic techniques for constructing non-supersymmetric solutions that exploit the integrability structure of supergravity theories. With such techniques at hand one can construct a variety of new solutions, in particular, new black hole solutions generalizing \cite{Elvang:2004xi, Compere:2010fm} and new fuzzball solutions
generalizing
\cite{Jejjala:2005yu, Bena:2009qv}.

The structure of this article is as follows. In sections~\ref{sec:BM} and~\ref{sec:BZ} we
introduce the linear systems of Breitenlohner--Maison and of Belinski--Zakharov
in Ehlers form and elucidate their interrelation. In section~\ref{sec:solitons} we study
meromorphic generating functions and solve the related linear systems
algebraically. Section~\ref{sec:KN} gives the explicit example of the Kerr--NUT solution in both cases.
We conclude in section \ref{conclusions}. Certain  technical computations have been
relegated to the appendices.

\section{Breitenlohner--Maison linear system}
\label{sec:BM}

In this section we review the linear system of Breitenlohner and Maison (BM)~\cite{Breitenlohner:1986um,Julia:1996nu,Maison,Nicolai:2001cc}. This linear system arises from considering a $D=3$ gravity-matter
system with target being a symmetric space G$_\EE$/K$_\EE$ where G$_\EE$ is the
global symmetry group that we refer to as Ehlers symmetry. Our conventions are
such that the three-dimensional space has signature $+++$ and we consider the space of solutions admitting an additional (axial) symmetry so that the theory becomes effectively two-dimensional. This two-dimensional system can be shown to be integrable.

\subsection{$D=3$ model with $\mathrm{G}_\EE/\mathrm{K}_\EE$ matter}

Consider a three-dimensional Euclidean gravity-matter system with action
\begin{align}
\label{gen3d}
S_\EE= \int d^3x \sqrt{-g} \left( R - \langle P_{\EE,\mu} | P_{\EE,\nu} \rangle g^{\mu\nu}\right),
\end{align}
where $\langle\cdot | \cdot \rangle$ is the symmetric invariant bilinear form on
the Lie algebra $\lag_\EE$ of the real Lie group $\mathrm{G}_\EE$. The group $\mathrm{K}_\EE$ is a
subgroup of $\mathrm{G}_\EE$ with dimension equal to that of the maximal compact
subgroup and the coset space $\mathrm{G}_\EE/\mathrm{K}_\EE$ is a (pseudo-)Riemannian symmetric
space. The subgroup $\mathrm{K}_\EE$ is fixed by some involutive automorphism $\tau$ on $\mathrm{G}_\EE$~\cite{Knapp}. This induces an involution $\theta$ on the Lie algebra (equal to the Cartan involution when $\mathrm{K}_\EE$ is maximally compact) which splits $\lag_\EE=\lak_\EE\oplus\lap_\EE$. Let $V_\EE\in \mathrm{G}_\EE/\mathrm{K}_\EE$ be a coset representative (e.g. in Borel gauge according to the Iwasawa theorem in a patch where it applies); then we decompose
\begin{align}
\partial_\mu V_\EE V_\EE^{-1}   = P_{\EE,\mu} + Q_{\EE,\mu}
\end{align}
with
\begin{subequations}
\label{qpdef}
\begin{align}
Q_{\EE,\mu} &%
=\frac12 \left(  \partial_\mu V_\EE V_\EE^{-1}- ( \partial_\mu V_\EE V_\EE^{-1})^\#\right),\\
P_{\EE,\mu} &%
= \frac12 \left(  \partial_\mu V_\EE V_\EE^{-1}+( \partial_\mu V_\EE V_\EE^{-1})^\#\right),
\end{align}
\end{subequations}
where we have defined the `generalized transpose' $x^\#= -\theta(x)$ on the Lie algebra. (For $\mathfrak{sl}(n,\reals)$ it is the standard transpose when $\lak=\mathfrak{so}(n)$.)  The two components in the splitting (\ref{qpdef}) satisfy
\begin{align}
Q_\EE^\# = -Q_\EE, \qquad \qquad P_\EE^\# = P_\EE.
\end{align}
The map $\#$ is an anti-involution and its group version will be denoted by the same symbol and is also an anti-involution: For $g,h\in \mathrm{G}_\EE$ one has $g^\# = \tau(g^{-1})=\tau(g)^{-1}$ and $(gh)^\#=h^\#g^\#$.

The symmetry transformations acting on $V_\EE$ are
\begin{align}
\label{GKsym}
V_\EE(x)\to k_\EE (x)V_\EE(x) g_\EE
\end{align}
with constant $g_\EE\in\mathrm{G}_\EE$ (global transformations) and varying $k_\EE \in \mathrm{K}_\EE$ (gauge transformations). Under these transformations, the Lie algebra valued quantities transform as
\begin{subequations}
\begin{align}
Q_{\EE,\mu} &\to k_\EE Q_{\EE,\mu} k_\EE^{-1} + \partial_\mu k_\EE k^{-1}_\EE ,\\
P_{\EE,\mu} &\to k_\EE Q_{\EE,\mu} k_\EE^{-1},
\end{align}
\end{subequations}
i.e., $Q_\EE$ as a $\mathrm{K}_\EE$-connection and $P_\EE$ transforms $\mathrm{K}_\EE$-covariantly.

A useful quantity associated with $V_\EE\in \mathrm{G}_\EE/\mathrm{K}_\EE$ is the `monodromy matrix'
\begin{align}
\label{MEE}
M_\EE = V_\EE^\# V_\EE
\end{align}
that transforms as $M_\EE\to g_\EE^\# M_\EE g_\EE$ under (\ref{GKsym}) since $k_\EE^\#k_\EE=\id$. It is hence insensitive to the $\mathrm{K}_\EE$-gauge chosen for the coset representative $V_\EE$ and only transforms under the global $\mathrm{G}_\EE$ transformation.

The $D=3$ equations of motion derived from (\ref{gen3d}) are
\begin{subequations}
\label{3deoms}
\begin{align}
\label{3deinst}
R_{\mu\nu} - \langle P_{\EE,\mu} | P_{\EE,\nu} \rangle &=0,&\\
\label{3dmatter}
\partial_\mu\left(\sqrt{-g}g^{\mu\nu} P_{\EE,\nu}\right) -\sqrt{-g} g^{\mu\nu} \lb Q_{\EE,\mu},P_{\EE,\nu}\rb  &=0.&
\end{align}
\end{subequations}
For convenience, we introduce the $K_\EE$-covariant derivative
\begin{align}
D_{\EE,\mu} = \partial_\mu - \lb Q_{\EE,\mu},\cdot\rb
\end{align}
in terms of which (\ref{3dmatter}) becomes
\begin{align}
D_{\EE,\mu} (\sqrt{-g} P_\EE^\mu) = 0.
\end{align}

\subsection{Reduction to $D=2$}

If the system admits an axial isometry $\partial_\phi$ we reduce the metric according to
\begin{align}
\label{2dredmet}
ds_3^2 = \cf_\EE^2 ds_2^2 + \rho^2 d\phi^2.
\end{align}
The function $\cf_\EE$ will be referred to as the conformal factor of the effective
two-dimensional metric. We label the two-dimensional coordinates as $x^m$. The two-dimensional metric $ds_2^2$ is assumed to be flat by appropriate choice of coordinates.
Note that there is no Kaluza--Klein vector $A_m$ of the type $(d\phi+A_m dx^m)^2$ in (\ref{2dredmet}) since it carries no degrees of freedom and can be set to zero without loss of generality.

With this ansatz, the equations of motion (\ref{3deoms}) then imply
\begin{align}
\square \rho &= 0,
\end{align}
which we solve by choosing Weyl canonical coordinates $x^m=(\rho,z)$ on
the flat two-dimensional space so that $ds_2^2=d\rho^2+dz^2$. We let $z$ be the conjugate variable to $\rho$ such that $\partial_\rho= \epsilon^{\rho z} \partial_z = \star_2 \partial_z$. It is often useful to combine the two real variables into a single complex variable (and its complex conjugate) which we denote by
\begin{align}
x^{\pm} = \frac12\left(z\mp i \rho\right)
\label{lightcone}
\end{align}
in analogy with light-cone coordinates that would arise if the two-dimensional base was Minkowskian as for colliding plane wave solutions. Note that (\ref{lightcone}) implies $\partial_\pm\rho = \pm i$ and $\star_2 \partial_{\pm} = \pm i \partial_{\pm}$.

With this choice the remaining equations are equivalent to
\begin{subequations}
\label{2deoms}
\begin{align}
\label{confconstraint}
\pm i \cf_\EE^{-1}  \partial_\pm\cf_\EE &= \frac\rho2 \langle P_{\EE,\pm}| P_{\EE,\pm}\rangle,\\
\label{2dmatter}
D_m(\rho\, P_\EE^m) &=0.
\end{align}
\end{subequations}
The first equation is a constraint and yields the conformal factor by a single integration and is therefore of secondary interest. The main equation of interest in this paper is the last equation (\ref{2dmatter}). A key result of~\cite{Breitenlohner:1986um} is that this equation is integrable and has an underlying symmetry structure associated with the affine extension of $\mathrm{G}_\EE$.

\subsection{BM linear system in general}

Consider the generalization of $V_\EE \in \mathrm{G}_\EE/\mathrm{K}_\EE$ to also depend on some spectral parameter $t$ (we suppress the space-time dependence) $V_\EE \longrightarrow \cV_\EE(t)$, where
\begin{align}
\label{calVtriangular}
\cV_\EE(t) = V^{(0)}_\EE + t V^{(1)}_\EE +\frac12 t^2V^{(2)}_\EE +\ldots.
\end{align}
importantly is regular in $t$ around $t=0$ and the limit $t\to 0$ gives back the original $V_\EE$:
\begin{align}
\lim_{t\to 0} \cV_\EE(t)= \cV_\EE(0) = V^{(0)}_\EE := V_\EE.
\end{align}
Consider the linear system~\cite{Breitenlohner:1986um,Nicolai:2001cc}
\begin{align}
\label{LSgen}
\partial_\pm \cV_\EE \cV_\EE^{-1}=  \frac{1\mp i t}{1\pm i t} P_{\EE,\pm}+
Q_{\EE,\pm},
\end{align}
where $Q_\EE$ and $P_\EE$ are independent of the spectral parameter $t$ and are defined  in terms of the $t$-independent $V_\EE$ as in (\ref{qpdef}).\footnote{In a general coordinate system, the linear system takes the form
\begin{align}
\partial_m \cV_\EE \cV^{-1}_\EE = Q_{\EE,m} + \frac{1-t^2}{1+t^2} P_{\EE,m} -\frac{2t}{1+t^2} \epsilon_{mn} P_\EE^n.\nn
\end{align}}
The integrability condition for (\ref{LSgen}) is equivalent to the equation of motion (\ref{2dmatter}) if and only if the spectral parameter $t$ satisfies the differential equation
\begin{align}
t^{-1} \partial_\pm t =  \frac{1\mp i t}{1\pm i t} \rho^{-1}\partial_\pm \rho.
\label{partialt}
\end{align}
This differential equation can be integrated~\cite{Breitenlohner:1986um} to an equation for $t$ which we write in the more conventional form in terms of the Weyl coordinates $(\rho,z)$ (cf. (\ref{lightcone}))
\begin{align}
\label{RSurface}
t^2 -\frac{2t}\rho (z-w) -1 =0,
\end{align}
where $w$ is an integration constant. This quadratic equation has two solution branches
\begin{align}
\label{BMspecs}
t_{\pm}= \frac1\rho\left[ (z-w) \pm \sqrt{(z-w)^2+\rho^2}\right].
\end{align}
Equation (\ref{RSurface}) defines a two-sheeted Riemann surface over the
two-dimensional flat base. We take the solution with the $+$ sign to be the
physical sheet and when we write $t$ we always mean $t_+$ unless indicated otherwise. We will refer to $t$ as the {\em space-time dependent spectral parameter} and to $w$ as the {\em constant spectral parameter}.\footnote{When written in terms of the `light-cone' coordinates (\ref{lightcone}), equation (\ref{BMspecs}) becomes
\begin{align}
t_{\pm} = - i \left[\frac{\sqrt{w-2x^+}\mp \sqrt{w-2x^-}}{\sqrt{w-2x^+}\pm \sqrt{w-2x^-}}\right].\nn
\end{align}
}

We will refer to (\ref{LSgen}) as the {\em BM linear system (in Ehlers form)} and we have just reviewed how its integrability condition gives rise to the equations of motion of the $D=3$ gravity-matter system (\ref{gen3d}) in the presence of a Killing isometry. This establishes the integrability of the equation (\ref{2dmatter}); the conformal factor $\cf_\EE$ can then be obtained by integrating (\ref{confconstraint})~\cite{Breitenlohner:1986um}. The linear system (\ref{LSgen}) is vastly underdetermined since it represents two differential equations for a function of three variables. There is an infinity of integration constants associated with this system.

Besides giving the integrability of (\ref{2dmatter}) the BM linear system also serves to unveil the group theory underlying the system. The original
$t$-independent $V_\EE \in \mathrm{G}_\EE/\mathrm{K}_\EE$ transformed under
global Ehlers transformations $g_\EE\in \mathrm{G}_\EE$ as $V_\EE \to k_\EE V_\EE g_\EE$ (cf. \eqref{GKsym}), where $k_\EE\in \mathrm{K}_\EE$ is the usual local compensator required to restore a chosen gauge for the coset representative. The presence of the spectral parameter now suggests to enlarge the set of global symmetry transformations by allowing $g_\EE$ to depend on the constant spectral parameter $w$:
\begin{align}
\label{afftrm}
\cV_\EE(t) \to k_\EE(t) \cV_\EE(t) g_\EE(w).
\end{align}
As indicated, the compensator is now also $t$-dependent as it has to be chosen such that the transformed $\cV_\EE(t)$ is regular around $t=0$ as in (\ref{calVtriangular}). This enlarged set of global transformations consists therefore of functions $g_\EE(w)$, i.e., maps of the type $\cx\to \mathrm{G}_\EE$, where we impose that $g_\EE(w)$ admits an expansion around $w= \infty$ in order to remain expandable as in (\ref{calVtriangular}).
These maps include transformations from $S^1\subset \cx$ into $\mathrm{G}_\EE$ and (under additional regularity assumptions) will lead to the loop group $\hat{G}_\EE$ associated with the Ehlers group $G_\EE$. Therefore the group underlying the integrability in $D=2$ includes the infinite-dimensional loop group; in fact the extension to the full affine group is active~\cite{Julia1981,Breitenlohner:1986um} where the central extension acts on the conformal factor $\cf_\EE$ (see below). We note that besides the affine group one can also define the action of the (centerless) Virasoro algebra which arises from arbitrary reparametrisations of the constant spectral parameter~\cite{Hou:1987xq,Maison:1988zx,Julia:1996nu}. Together with the infinitesimal affine transformations one obtains a semi-direct product in the standard way. We will not use the Virasoro symmetry in this paper.

The involution $\#$ extends to functions $\cV_\EE(t)$ by
\begin{align}
\label{hashextend}
(\cV_\EE(t))^\# = \cV_\EE^\#\left(-\frac1t\right).
\end{align}
One can use $\#$ to split the loop algebra into an invariant and an anti-invariant part (generalizing $\lap_\EE$ and $\lak_\EE$ above). Now, it is important that the right hand side of (\ref{LSgen}) is anti-invariant under (the Lie algebra version of) $\#$ and therefore belongs to the `compact' subalgebra of the affine algebra based on $\lag_\EE$. This anti-invariance implies that $((\cV_\EE(t))^\#)^{-1}$ is a solution to the linear system if $\cV_\EE(t)$ is a solution. In general, the two solutions related by this involutive mapping will be different. The mapping implies that the monodromy matrix
\begin{align}
\label{monG}
\cM_\EE = \cM_\EE(w) = \left(\cV_\EE(t)\right)^\# \cV_\EE(t) = \cV_\EE^\#\left(-\frac1t\right) \cV_\EE(t)
\end{align}
is independent of the space-time coordinates $x^m$ and therefore a function of $w$ alone. The matrix $\cM_\EE(w)$ is invariant under the application of $\#$ and simultaneously exchanging $t\to -1/t$. We note that this is evident from (\ref{RSurface}) which implies that $w(t,x)$ is invariant under $t\to-1/t$.

Constructing new solutions of the linear system by means of the Geroch symmetry proceeds along the following chain of steps\footnote{We present here the solution generating method based on the monodromy matrix $\cM_\EE(w)$. Alternatively, one could work at the level of the generating function $\cV_\EE(t)$ and the transformation (\ref{afftrm}); however, the step of finding the compensator $k_\EE(t)$ in (\ref{afftrm}) is typically very hard. }
\begin{align}
\label{BMsol}
V_\EE \to \cV_\EE(t) \to \cM_\EE(w) \to \cM^g_\EE(w) \to \cV^g_\EE(t) \to V^g_\EE,
\end{align}
where we introduced the notation
\begin{align}
\label{montrm}
\cM_\EE(w) \to \cM_\EE^g(w):=g_\EE^\#(w) \cM_\EE(w) g_\EE(w)
\end{align}
for the transformed solution. The individual steps in (\ref{BMsol}) starting from a given seed solution $V_\EE$ are: $(i)$ find a corresponding generating function $\cV_\EE(t)$ that solves the linear system (\ref{LSgen}), $(ii)$ compute the associated monodromy, $(iii)$ transform the monodromy under a global transformation $g(w)$ as in (\ref{montrm}), $(iv)$ factorize the new monodromy into a new generating function $\cV^g_\EE(t)$ and $(v)$ take the limit $t\to 0$ to find the new solution.

For practical purposes, the main difficulty resides in step $(iv)$ in factorizing the transformed $\cM_\EE^g(w)$ as
\begin{align}
\label{RHgen}
\cM_\EE^g(w) = (\cV_\EE^g(t))^\# \cV_\EE^g(t)
\end{align}
with the new $\cV^g_\EE(t)$ having an expansion as in (\ref{calVtriangular}). This is a Riemann--Hilbert problem~\cite{Breitenlohner:1986um} whose solution is in general hard to obtain. In the particular case of meromorphic $\cM^g_\EE(w)$ with single poles in $w$ of certain simple type one can reduce the problem to a set of linear algebraic equations. This is the case of soliton charging transformations that will be discussed further in section~\ref{sec:solitons}. Once the new $\cV^g_\EE(t)$ has been obtained, one can recover the solution to the gravity-matter system (\ref{2deoms}) by taking the limit $t\to 0$ and obtain $V^g_\EE\in \mathrm{G}_\EE/\mathrm{K}_\EE$ that characterises the physical fields.

Besides the knowledge of the coset `scalars' $V^g_\EE\in \mathrm{G}_\EE/\mathrm{K}_\EE$ one also requires the new conformal factor $\cf_\EE^g$ in (\ref{2deoms}). This can be obtained from a simple integration of (\ref{confconstraint}) but it also follows from group theoretic properties using the central extension. This is discussed in detail in~\cite{Breitenlohner:1986um} to which we refer for the general expression. In section~\ref{sec:solitons} we will present the formula in the case of soliton transformations.

We note that a trivial solution of the equations (\ref{LSgen}) and (\ref{confconstraint}) is given by
\begin{align}
\label{BMflat}
\cV_\EE(t) = \id\quad\textrm{and}\quad \cf_\EE =1.
\end{align}
This solution will be referred to as  flat space as it corresponds to the Minkowski vacuum in the four-dimensional case.

\section{Belinski--Zakharov linear system}
\label{sec:BZ}

In this section we present the linear system used by Belinski and Zakharov (BZ)~\cite{BZ,BV}. We will not present it in the standard form which uses what was called the Matzner--Misner formulation in the introduction. Rather, we will use the Ehlers description to make contact with the discussion in the preceding section. (The Matzner--Misner version and its relation to BM is discussed in appendix~\ref{app:BZMM}.)

\subsection{BZ Ehlers linear system}

Equation \eqref{2dmatter} for the $\mathrm{G}_\EE/\mathrm{K}_\EE$ coset fields admits an alternative Lax pair that can be written as
\begin{align}
\label{BZLaxM}
D_1 \Psi_{\EE} = \frac{\rho V-\lambda U}{\lambda^2+\rho^2}\Psi_{\EE},\qquad \qquad
D_2 \Psi_{\EE} = \frac{\rho U +\lambda V}{\lambda^2+\rho^2}\Psi_{\EE},
\end{align}
where $\lambda$ is the (space-time independent) spectral parameter of BZ and
$\Psi_{\EE}(\lambda,\rho,z)$ is the generating function such that the matrix
$M_\EE=V^\#_\EE V_\EE$ of (\ref{MEE}) is recovered for $\lambda=0$:
\begin{align}
 M_\EE(\rho,z)=\Psi_{\EE}(0,\rho,z).
\end{align}
The matrices $U,V$ are defined as
$ U=\rho\partial_\rho M_\EE M_\EE^{-1}$, $V=\rho\partial_z M_\EE M_\EE^{-1}$,
and the differential operators $D_1,D_2$ are
\begin{align}
\label{BZD1D2}
D_1 = \partial_z - \frac{2\lambda^2}{\lambda^2+\rho^2}\partial_\lambda,\qquad \qquad
D_2 = \partial_\rho + \frac{2\lambda \rho}{\lambda^2+\rho^2}\partial_\lambda.
\end{align}
The operators $D_1$ and $D_2$ commute and the associated integrability condition of the linear system (\ref{BZLaxM}) is equivalent to the desired non-linear equation (\ref{2dmatter}).

Solutions of the BZ linear system (\ref{BZLaxM}) are constructed using the inverse scattering method~\cite{BZ}. One starts from a `seed'
$\Psi_{\EE,0}$, that is `dressed' to obtain a new solution $\Psi_{\EE}$ through
\begin{align}
 \Psi_\EE(\lambda)=\chi(\lambda) \Psi_{\EE,0}(\lambda)
\end{align}
where $\chi$ is called the dressing matrix and it depends on the spectral parameter $\lambda$.
The seed $\Psi_{\EE,0}$ corresponds to a solution of (\ref{BZLaxM}) for a seed solution $M_{\EE,0}$.
We can take it to be the identity matrix $\Psi_{\EE,0}=\id$, which corresponds to taking the seed solution to be flat space\footnote{Note that this differs from the more common (and complicated) expression for flat space in the Matzner--Misner form~\cite{BZ,BV}.}.
In order for the `dressed' $\Psi_{\EE}$ to also solve the linear system (\ref{BZLaxM}) the dressing matrix has to satisfy its own linear system
\begin{align}
\label{LSchi}
 D_1 \chi = \frac{\rho V - \lambda U}{\lambda^2 + \rho^2} \chi - \chi
\frac{\rho V_0 - \lambda U_0}{\lambda^2 + \rho^2}, \qquad D_2 \chi = \frac{\rho
U + \lambda V}{\lambda^2 + \rho^2} \chi - \chi \frac{\rho U_0 + \lambda
V_0}{\lambda^2 + \rho^2}.
\end{align}
In addition, the matrix $\chi$ must satisfy further constraints in order to ensure that the new solution $M_\EE(\rho,z)=\Psi(0,\rho,z)$ is real, satisfies $M_\EE^\#=M_\EE$ and is a representative of the coset $\mathrm{G}_\EE/\mathrm{K}_\EE$~\cite{BZ,Figueras:2009mc,Shabnam}.

\subsection{Relation between the two linear systems}

Compared to the discussion of the BM linear system, the differential operators $D_1$ and $D_2$ of (\ref{BZD1D2}) can be demystified by thinking of $\lambda$ as space-time dependent, so that~\cite{Breitenlohner:1986um}
\begin{align}
 \label{BZdiffM}
D_1 = \partial_z = \partial_z|_{\lambda\,\text{fixed}} + \partial_z \lambda \partial_\lambda,\quad
D_2 = \partial_\rho = \partial_\rho|_{\lambda\,\text{fixed}} + \partial_\rho \lambda \partial_\lambda.
\end{align}
If  the spacetime dependence of the spectral parameter $\lambda$ is given by
\begin{align}
\label{lambdaeq}
\partial_z \lambda = -\frac{2\lambda^2}{\lambda^2+\rho^2},\quad
\partial_\rho \lambda = \frac{2\lambda\rho}{\lambda^2+\rho^2}
\end{align}
one recovers (\ref{BZD1D2}). The solution to (\ref{lambdaeq}) is given by
\begin{align}
\lambda(\rho,z) = (w-z)\mp \sqrt{(z-w)^2+\rho^2},
\end{align}
where $w$ is an integration constant. Comparing to  \eqref{BMspecs}, it follows that from this viewpoint the relation of the BZ spectral parameter $\lambda$ to $t$ in the BM approach~\cite{Breitenlohner:1986um,Figueras:2009mc} is\footnote{The sign in this relation in~\cite{Breitenlohner:1986um} appears incorrect.}
\begin{align}
\label{2specs}
 \lambda(\rho,z)=-\rho t(\rho,z).
\end{align}

The relation between the two generating functions $\cV_\EE$ of (\ref{LSgen}) and $\Psi_\EE$ of (\ref{BZLaxM}) is given by
\begin{align}
\Psi_\EE (\lambda,x)= V^\#_\EE(x) \cV_\EE(t,x) \label{relationBZBM}
\end{align}
where one also has to use (\ref{2specs}). Note that on the right-hand side we have once the spectral parameter independent $V_\EE(x)=\cV_\EE(0,x)$ and once the full $\cV_\EE(t,x)$. This obscures the action (\ref{afftrm}) of the full affine Geroch group since the transformation of $V_\EE(x)$ under affine elements is complicated.

In the following we restrict to $\mathrm{G}_\EE=\mathrm{SL}(n,\reals)$ for concreteness. In that case $M_\EE$ has to be a symmetric matrix. For other groups, there will be different conditions on some of the quantities introduced below.

If the matrices $M_\EE$ and $M_{\EE,0}$ obtained by the $\lambda\to 0$ limits of $\Psi_\EE$ and $\Psi_{\EE,0}$ are symmetric, then
\begin{align}
\label{trchi}
\chi'(\lambda) = M_\EE \chi^{T^{-1}}\left(-\frac{\rho^2}{\lambda}\right) M_{\EE,0}^{-1},
\end{align}
solves exactly the same linear system \eqref{LSchi} as $\chi$ \cite{BZ}. Given this observation,
one has that $\chi'(\lambda)$ is related to $\chi(\lambda)$ through some arbitrary matrix $C(w)$ via
\begin{align}
\label{chicondtemp}
\chi'(\lambda) \Psi_{\EE,0}
= \chi(\lambda) \Psi_{\EE,0}
C(w).
\end{align}
This reflects the fact that the linear system (\ref{LSchi}) is underdetermined and $C(w)$ corresponds to a gauge freedom of (\ref{BZLaxM}).
However, Belinski and Zakharov demand
\begin{align}
\label{chicond}
\chi'(\lambda)=\chi(\lambda),
\end{align}
which corresponds to fixing the gauge freedom of (\ref{chicondtemp}). In addition, they do \emph{not} require $\chi(\lambda)$ to satisfy the coset constraint $\det \chi(\lambda) = 1$. Since $\det\chi \neq 1$ one has that the new matrix $M_\EE$ does not have unit determinant and so does not represent a physical configuration. To obtain the `physical' matrix  $M^{\mathrm{(phys)}}_\EE$ that fullfills the determinant condition, one rescales the
matrix $M_\EE$ appropriately.

Condition \eqref{chicond} automatically ensures that the final $M_\EE$ is symmetric, but it is a rather strong assumption. Relation (\ref{chicond}) is a central equation in the BZ inverse scattering framework and it fixes an infinite ambiguity in the dressing matrix that corresponds roughly to the Borel part of the Geroch group. In other words, demanding (\ref{chicond}) in the BM framework amounts to choosing finely tuned integration
constants for all the dual potentials $V_\EE^{(n)}$ with $n \ge 1$ introduced through \eqref{calVtriangular}. The transformation (\ref{trchi}) is similar to the one discussed below (\ref{hashextend}) in the BM framework. There one normally does not fix this freedom and so there is no direct analogue of (\ref{chicond})
in the BM approach.

Due to the complications of $\det\chi\neq 1$ and the issues mentioned around \eqref{relationBZBM}, it is impractical to find a satisfactory embedding of the full BZ solution generating technique in the Geroch group\footnote{Naively one might conclude
from  \eqref{chicondtemp}--\eqref{chicond} that it simply corresponds to taking $C(w) = \id$
from the Geroch group point of view. However, this interpretation is not correct. This is because the dressed BZ matrix $\Psi_\EE(\lambda)= \chi(\lambda) \Psi_{\EE,0}$
does not directly give the physical matrix $M^{\mathrm{(phys)}}_\EE$. In order to have an interpretation of $C(w)$ in the Geroch group, one first needs to construct $\chi^{\mathrm{(phys)}}(\lambda)$. A procedure to do this was
suggested in \cite{Cosgrove:1980fx}.  Requiring something like $\chi^{\mathrm{(phys)}}{}'(\lambda) = \chi^{\mathrm{(phys)}}(\lambda)$ will indeed be more
amenable to the group theoretic
interpretation, but it is not the BZ technique.}. The best one can do is to find a representative relation between
the
BZ-generating function $\Psi_\EE(\lambda)$ and the group-theoretic BM generating
function $\cV_\EE(t)$. This relation is precisely equation \eqref{relationBZBM} for the Ehlers coset and is obtained in Appendix \ref{app:BZMM} for the Matzner--Misner coset, see \eqref{relationBZBMMM}.

\subsection{Solitonic solutions}
\label{BZsols}

So-called {\em solitonic solutions} for the matrix $M_\EE$ correspond to a dressing matrix
$\chi(\lambda)$ with simple poles in the complex $\lambda$-plane. The general $N$-soliton solution is obtained by dressing the seed solution with a matrix $\chi$ of the form
\begin{align}
\label{nsolchi}
 \chi=\id +\displaystyle\sum_{k=1}^{N}\frac{R_k}{\lambda-\mu_k}.
\end{align}
The matrices $R_k$ and the pole trajectories $\mu_k$ are functions of $\rho,z$ only. For each soliton, there exist two possible solutions for the pole trajectory $\mu_k$
\begin{align}
\label{BZpolepos}
 \mu_k=-\left(z-w_k\right)\pm \sqrt{\left(z-w_k\right)^2 +\rho^2},
\end{align}
where the parameters $w_k$ may generally be complex but for the examples considered here we will take them to be real. The pole trajectories with a ``$+$'' sign are
referred to as solitons
and the ones with a ``$-$'' sign as antisolitons.

In order to construct the $N$-soliton dressing matrix, one needs to parametrise the residue matrices $R_k$. Here, one has the freedom of introducing certain arbitrary {\em constant} parameters $m_{0b}^{(k)}$ (with $b=1,\ldots,n$ when $\Psi$ is represented as an $n\times n$-matrix) for each soliton $\mu_k$
as follows\footnote{The normalization of each of the vectors $m_{0b}^{(k)}$  is arbitrary. Rescaling them by arbitrary constants does not change any of the final expressions.\label{BZvector_norm}}. Defining
\begin{align}
\label{BZvectors}
 m_a^{(k)}= m_{0b}^{(k)}\left[\Psi_{\EE,0}^{-1}(\mu_k,\rho,z)\right]_{ba},
\end{align}
and the symmetric matrix $\Gamma_{\mathrm{BZ}}$ as
\begin{align}
\label{GammaM}
(\Gamma_{\mathrm{BZ}})_{kl}=\frac{m_a^{(k)}(M_{\EE,0})_{ab}m_b^{(l)}}{\rho^2+\mu_k\mu_l},
\end{align}
the elements of the residue matrices $R_k$ are given by
\begin{align}
\label{ResM}
\left(R_k\right)_{ab}=m_a^{(k)}\displaystyle\sum_{l=1}^{N}\frac{\left(\Gamma_{\mathrm{BZ}}^{-1}\right)_{lk} m_c^{(l)}\left(M_{\EE,0}\right)_{cb}}{\mu_l}.
\end{align}
The new matrix $M_\EE(\rho,z)=\chi(0,\rho,z)\Psi_{\EE,0}(0,\rho,z)$  now reads
\begin{align}
\label{newsolM}
 \left(M_\EE\right)_{ab}=\left(M_{\EE,0}\right)_{ab}-\displaystyle\sum_{k,l=1}^{N}\frac{\left(M_{\EE,0}\right)_{ac}m_c^{(k)}\left(\Gamma_{\mathrm{BZ}}^{-1}\right)_{kl}m_d^{(l)}\left(M_{\EE,0}\right)_{db}}{\mu_k \mu_l}.
\end{align}
The symmetry of this expression is ensured by (\ref{chicond}).
A problem that arises at this stage is that possibly the new matrix $M_\EE$  does not
satisfy the coset constraint $\det M_\EE=1$, i.e. is not an element of
the group $\mathrm{SL}(n,\mathbb{R})_\EE$. In fact the determinant of the new matrix is given by
\begin{align}
\label{detMun}
 \det M_\EE=(-1)^N \rho^{2N}\left(\displaystyle\prod_{k=1}^{N}\mu_k^{-2}\right)\text{det}M_{\EE,0}.
\end{align}
In order to obtain an $N$-soliton solution that remains in the group SL$(n,\mathbb{R})_\EE$, the new matrix $M_\EE$ must be multiplied by an overall factor\footnote{This formula differs slightly from the standard expression in~\cite{BZ} since we are working in the Ehlers description.}
\begin{align}
\label{Mphys}
M^{\mathrm{(phys)}}_\EE=\pm\left(\frac{1}{\pm\text{det}M_\EE}\right)^{\frac{1}{n}} M_\EE.
\end{align}
The overall sign in
this expression should be chosen in order to
ensure the right metric signature. Thus obtained $M^{\mathrm{(phys)}}_\EE$ fulfils the constraint $\text{det}M^{\mathrm{(phys)}}_\EE=1$. Finally, following the discussion in \cite{BZ}, the conformal factor for the dressed solution can also be obtained. We find for $\mathrm{SL}(2,\reals)_\EE$
\begin{align}
\label{conformal_factorE}
(\cf^{\mathrm{(phys)}}_\EE)^2 = k_\mathrm{BZ} \cdot \rho^{ N-\frac{N^2}{2}} \cdot \left( \prod_{k=1}^{N} \mu_k\right)^N  \cdot \left[\prod_{k, l = 1, \; k > l}^{N}(\mu_k - \mu_l)^2\right]^{-1} \cdot \mbox{det} \Gamma_\mathrm{BZ} \cdot\cf_{\EE,0}^2,
\end{align}
where $k_\mathrm{BZ}$ is an arbitrary numerical constant. For $\mathrm{SL}(n,\reals)_\EE$ a similar but more complicated expression can also be written \cite{BV}. However, note that for $n>2$ the rescaling (\ref{Mphys}) introduces fractional powers of $\rho$ from (\ref{detMun}) that typically lead to singular solutions. For this reason it is more useful to employ the so-called Pomeransky trick \cite{Pomeransky} for $n>2$. In this approach one can write a general expression for the conformal factor valid for $n\ge 2$ \cite{Pomeransky, Figueras:2009mc}.

\section{BM Soliton transformations}
\label{sec:solitons}

In this section we present an algebraic method of generating new solutions of the BM linear
system (\ref{LSgen}) from a given seed solution. Our discussion closely follows that of~\cite{BMunpub}, see also~\cite{NicolaiLectures}.

The method makes use of the constant group element $g_{\mathrm{E}}(w)$ of the Geroch group. We take the seed
solution to be flat space (\ref{BMflat}) since it is believed that the Geroch group action is transitive on the space of solutions and all solutions are related to flat space~\cite{EH,Breitenlohner:1986um,Bernard:1997et}. As mentioned around (\ref{RHgen}), the action of the Geroch group generally leads to a matrix valued Riemann--Hilbert problem. In the case when the matrix functions to be factorized
are meromorphic in the spectral parameter $w$, the problem can be solved
algebraically. This is the case that we focus on and we term it the solitonic case.
There are a number of (formal) similarities and at the same
time a number of differences (in details) with the procedure of Belinski and
Zakharov~\cite{BZ} that we briefly reviewed in section~\ref{BZsols}. In this section we also restrict ourselves to the
Ehlers $\mathrm{SL}(n,\reals)_\EE$ of $D=2+n$ vacuum gravity. In this case the generalized transpose $\#$
(at the group level) simply becomes the usual matrix transpose.

\subsection{Riemann--Hilbert problem}

In section \ref{sec:BM} we presented the construction of $\cM_\EE(w)$ starting with $V_\EE(x)$.
We start with $V_\EE(x)$, solve the linear system \eqref{LSgen} to find
$\cV_\EE(t, x)$, and then construct $\cM_\EE(w)$. We now ask, following the steps of (\ref{BMsol}), if we can reverse the
process and reconstruct $V_\EE(x)$ \emph{from} $\cM_\EE(w)$, i.e., we solve the
Riemann--Hilbert problem to factorize $\cM_\EE(w)$ as in~(\ref{monG}). It is not guaranteed that for a general $\cM_\EE(w)$ such a factorization exists. In fact, one can construct explicit examples where it does not exist. This is however a technical problem
that will not concern us here. Certain aspects of this have been studied in the literature, see e.g.~\cite{EH}.
We only work with those matrices $\cM_\EE(w)$ for which the Riemann--Hilbert problem admits a solution.

Let us start with a real symmetric unit determinant matrix $\cM_\EE(w)$ assuming suitable analyticity properties. In particular we assume
$\cM_\EE(\infty) = \id$. We wish to factorize it as
\begin{align}
\cM_\EE(w) = A_{-}^T(t,x) M_\EE(x) A_{+}(t,x),
\label{factor}
\end{align}
with $A_{-}(t,x) = A_{+} \left(-\frac{1}{t}, x \right)$ and $M_\EE(x)$ symmetric and real.
Moreover, we require
\begin{align}
 \det A_{\pm}(t,x) = 1.
\end{align}
Next, we factorize $M_\EE(x)$ as $M_\EE(x) = V_\EE^T(x)V_\EE(x)$
with a triangular matrix $V_\EE(x)$ to obtain
\begin{align}
\cV_\EE(t,x) = V_\EE(x)A_{+}(t,x).
 \label{finalcVE}
\end{align}

The factorization problem \eqref{factor} can be viewed in two different ways. $(i)$ We start with an appropriate $\cM_\EE(w)$ and solve for $\cV_\EE(t,x)$, $(ii)$ we start with a seed
$\cV_\EE(t,x)$ and act with an element $g_\EE(w)$ and attempt to determine the transformed $\cV^g_\EE(t,x)$. We take the first
viewpoint in what follows.\footnote{The second viewpoint was taken in~\cite{BMunpub}, however, some of their assumptions about pole structures seem too restrictive to make that method directly applicable to interesting solutions.}
It is, however, convenient
to have the second viewpoint in the back of one's mind and relate it to the first one by taking the seed to be flat space. At the level of equations this means
\begin{align}
\cV^g_\EE(t,x) = Z_+^g(t,x) \cV_\EE(t,x) = Z_+^g(t,x) \cdot \id = Z_+^g(t,x),
\end{align}
where $Z_+^g(t,x)$ is a triangular `dressing' matrix in the sense of (\ref{calVtriangular}) that is determined by the Geroch group element $g_\EE(w)$.
In terms of the monodromy matrix one similarly has
\begin{align}
\cM^g_\EE(w) = \cV^T_\EE\left(-\frac{1}{t},x\right) \left[Z_+^g \left(-\frac{1}{t},x\right)\right]^T  Z_+^g(t,x) \cV_\EE(t,x).
\end{align}
Introducing
\begin{align}
\cZ^g(t,x) = \left[Z_+^g\left(-\frac{1}{t},x\right)\right]^T  Z_+^g(t,x),
\end{align}
one also has
\begin{align}
\cZ^g(t,x) &= \left[\cV^T_\EE\left(-\frac{1}{t},x\right)\right]^{-1}
\cM_\EE^g(w) \left[\cV_\EE(t,x)\right]^{-1} \label{cZ} &\\
&= \id \cdot \cM_\EE^g(w) \cdot \id = \cM^g(w).&
\end{align}
When $\cV_\mathrm{E}(t,x) \neq \id$ one should take $\cZ^g(t,x)$ in \eqref{cZ}
to be the left hand side of equation \eqref{factor} and solve the corresponding factorization problem. In this paper we always work with flat space (\ref{BMflat}) as seed solution.
Consequently, for notational simplicity we drop the superscript $g$ from now on and just think of being given a monodromy $\cM_\EE(w)$ that needs to be factorized as in~(\ref{factor}).

\subsection{Multisoliton solutions}

The factorization problem can be solved algebraically when the matrix functions
to be factorized are meromorphic. We now present this factorization explicitly by adapting~\cite{BMunpub}.
We assume that  $\cM_\EE(w)$ has simple poles with residues of rank one. Since $\det \cM_\EE(w) = 1$ and $\cM_\EE(\infty) = \id$ the inverse matrix
$\cM^{-1}_\EE(w)$ also has poles at the same points with residues of rank one. If we have $N$ poles at points $w_{k}$ with $k = 1, 2, \ldots, N$ we can express $\cM_\EE(w)$ and
$\cM^{-1}_\EE(w)$ in the form
\begin{subequations}
\begin{align}
\label{Mexp}
\cM_\EE(w) &= \id + \sum_{k=1}^N \frac{A_k}{w-w_k}, &\\
\label{Minvexp}
\cM^{-1}_\EE(w) &= \id - \sum_{k=1}^N \frac{B_k}{w-w_k},&
\end{align}
\end{subequations}
with symmetric (since $\cM_\EE$ is symmetric) and {\em constant} residue matrices $A_k$ and $B_k$ of rank one.
This means that we can factorize these matrices as the outer product of vectors
\begin{align}
A_k = a_k \alpha_k a_k^T, \qquad \qquad B_k = b_k\beta_k b_k^T. \label{rank_one}
\end{align}
One could absorb $\alpha_k$ and $\beta_k$ in the definition of the constant vectors $a_k$ and $b_k$ respectively but we leave them explicit on purpose. They play a very important role: for a given set of $a_k$ and $b_k$
we can tune the $\alpha_k$ and $\beta_k$ appropriately to ensure that the matrices  $\cM_\EE(w)$ and $\cM^{-1}_\EE(w)$
have unit determinant. Despite this, there is an ambiguity in the factorization \eqref{rank_one} related to the normalization of the vectors $a_k$ and $b_k$.
Nothing must depend
on this choice of normalization. This will indeed be the case as will be apparent shortly.
At this stage we just remark that the constant vectors $a_k$ are the analog of the constant vectors $m_0^{(k)}$ of (\ref{BZvectors}) in the BZ method. The ambiguity related to the
factorization of rank one matrices in vectors is directly related to the ambiguity in the normalization of the vectors $m_0^{(k)}$ in the
BZ method (cf.~footnote \ref{BZvector_norm}).
As is well known in the BZ method, nothing depends on the overall
normalization of those vectors.

In order to factorize $\cM_\EE(w)$ as in (\ref{factor}) we have to change from the constant spectral parameter $w$ to the space-time dependent parameter $t$ through (\ref{BMspecs}), which implies
\begin{align}
\frac{1}{w-w_k} = \nu_k \left( \frac{t_k}{t-t_k}+ \frac{1}{1+t t_k}\right),
\end{align}
where the moving poles $t_k$ are determined by (\ref{BMspecs}) evaluated at $w_k$ with the plus sign and
\begin{align}
\nu_k = -\frac{2 t_k}{\rho\left(1 + t_k^2 \right)}.
\end{align}
As a function of $t$, $\cM_\mathrm{E}(t,x)$ has a total of $2N$ poles: $N$ poles
at $t=t_k$ and $N$ poles at $t=-1/t_k$. These have to be distributed among the factors $A_+$ and $A_-$ in (\ref{factor}). The analytic properties of the Riemann--Hilbert problem are such that
the poles at $t=-1/t_k$ come from $A_+(t)$ and those at $t=t_k$ from $A_-$. One therefore makes the ans\"atze~\cite{BMunpub}
\begin{subequations}
\begin{align}
\label{Vplus}
A_+(t) &= \id - \sum_{k=1}^N \frac{c_k t a_k^T}{1+t t_k},&\\
\label{Vplusinv}
A^{-1}_+(t) &= \id + \sum_{k=1}^N \frac{b_k t d^T_k}{1+tt_k},&
\end{align}
\end{subequations}
where the second equation arises in the factorization of $\cM^{-1}_\EE(w)$. These two equations introduce two new sets of vectors that we call $c_k$ and $d_k$.

The vectors $a_k$, $b_k$, $c_k$ and $d_k$ are not all independent and determining their relation amounts to solving the Riemann--Hilbert problem.
We first note that the constant matrices $A_k$ and $B_k$ are not independent since the two matrices $\cM_\EE(w)$ and $\cM^{-1}_\EE(w)$ are
inverses of each other. This determines the vectors $b_k$ from the $a_k$ up to scaling, a freedom that is reflected in the $\beta_k$ in (\ref{Minvexp}).

We can use the pole structure of $\cM_\EE(t,x) \cM^{-1}_\EE(t,x)$  to deduce some properties of the $a_k$ and $b_k$. To start with, the absence of double poles
at $t=-1/t_k$ in the product implies
that the vectors $a_k$ and $b_k$ are orthogonal:
\begin{align}
a_k^T b_k = 0 \qquad \qquad \text{for each $k$.}
\end{align}
From the absence of single poles at $t=-1/t_k$ in the product $\cM_\EE(t,x)\cM^{-1}_\EE(t,x)$ one deduces the relations
\begin{align}
a_k \alpha_k a_k^T \cA^{k} =  \cA_k b_k \beta_k b_k^T,
\label{nosinglepole}
\end{align}
with the definitions
\begin{align}
\cA^{k} =\left[ \cM_\EE^{-1}(t,x) + \frac{b_k \nu_k \beta_k
b_{k}^T }{1 + t t_k} \right]_{t=-\frac{1}{t_k}},
\qquad
\cA_{k} = \left[\cM_\EE(t,x) - \frac{a_k  \nu_k \alpha_k a_k^T}{1 + t t_k} \right]_{t=-\frac{1}{t_k}}.
\end{align}
Equation \eqref{nosinglepole} is satisfied if there exist  $\gamma_k$ such that
\begin{align}
a_k^T \cA^{k} = \gamma_k \nu_k \beta_k b_k^T \qquad \mbox{and} \qquad
\cA_{k} b_k = \gamma_k \alpha_k \nu_k a_k. \label{nosinglepole2}
\end{align}
Then both sides of \eqref{nosinglepole} are equal to $ \alpha_k \nu_k \beta_k  \gamma_k (a_k b_k^T)$.
We note that the $\gamma_k$ defined uniquely by (\ref{nosinglepole2}) depend on space-time.

The next non-trivial step is to determine the vectors $c_k$. With the ansatz
\eqref{Vplus}  for $A_+(t)$,
the requirement that $A_+(t)\cM_\EE^{-1}(t,x)$ have no poles at $t=-1/t_k$
gives the vector equation
\begin{align}
c_k \Gamma_{kl} = b_l, \label{solc}
\end{align}
where $\Gamma_{kl}$ is an $N \times N$ matrix with elements
\begin{align}
\label{Gammadef}
\Gamma_{kl} = \left\{ \begin{array}{ll}
 \frac{\gamma_k}{t_k} &\mbox{\qquad for \qquad $k=l$} \\
  \frac{1}{t_k-t_l} a_k^T b_l &\mbox{\qquad for \qquad $k \neq l$.}
       \end{array} \right.
\end{align}
Solving equation (\ref{solc}) for the $c_k$ we obtain the
matrix $A_+(t)$. A similar argument is used to construct $A_+^{-1}(t)$ in (\ref{Vplusinv}).
One finds the equation for the vectors $d_k$ to be $\Gamma_{kl} d_l  = a_k$. Solving this equation for the $d_k$ we can readily construct the
monodromy matrix $M_\EE$ by taking the limit $t\to\infty$ in (\ref{factor}) and using that $\cM_\EE(\infty)=M_\EE A_+(\infty)=\id$. The result is
\begin{align}
M_\EE = A^{-1}_+(\infty) = \id + \sum_{k,l=1}^{N} b_k t_k^{-1} (\Gamma^{-1})_{kl} a_l^T.
\label{final2}
\end{align}
Finally we factorize $M_\EE = V_\EE^T V_\EE$ and obtain the space-time fields. If needed, we can also construct explicitly the generating
function $\cV_\EE(t,x)$ from equation \eqref{finalcVE}.

At this stage it is instructive to investigate how the final answers
\eqref{Vplus} (in conjunction with \eqref{solc}) and \eqref{final2} are
insensitive to the ambiguity related to the rescaling of the vectors.
Note that if we rescale the vectors $a_k$ to $r_k a_k$ and
$b_k$ to $s_k b_k$, then we must rescale $\alpha_k$ and $\beta_k$ as $r_k^{-2}\alpha_k$ and $s_k^{-2}\beta_k$ respectively in order to
preserve the products \eqref{rank_one}.  This then means that the $\gamma_k$ scale as $r_k s_k \gamma_k$.
 It then immediately follows that the matrix $\Gamma_{kl}$
scales as $r_k s_l \Gamma_{kl}$. The inverse matrix $(\Gamma^{-1})_{kl}$ naturally scales with the inverse factor $r^{-1}_k s^{-1}_l$ and hence we see that the
final answers \eqref{Vplus} (in conjunction with \eqref{solc}) and
\eqref{final2} are insensitive to such rescalings.

\subsection{Conformal factor}

For solitonic solutions of the previous subsection the conformal factor
$\cf_\EE$ can also be obtained in a closed form. The final result is
\begin{align}
\cf_\EE^2 = k_\mathrm{BM} \cdot \prod_{k=1}^{N} (t_k \nu_k) \cdot \det \Gamma,
\label{conformal_factor}
\end{align}
where $k_\mathrm{BM}$ is an arbitrary numerical constant. This constant needs to be chosen appropriately in order to ensure certain physical properties (say, asymptotic flatness)
of the final space-time. A derivation of expression \eqref{conformal_factor} is given in appendix \ref{conformal_factor:APP}.

\section{Example: Kerr--NUT solution}
\label{sec:KN}

In this section we present a concrete implementation of the method of section \ref{sec:solitons} by constructing the Kerr--NUT metric. The construction illustrates
all the steps of the BM solitonic method. More complicated examples are certainly doable; we leave such a line of investigation for the future.

\subsection{Construction of general 2-soliton solution}

The main difficulty in constructing the general multi-soliton solutions using the BM group theoretic approach lies in finding meromorphic matrices $\cM_\EE(w)$ that satisfy the coset constraints.
The analog of this problem does not arise in the approach of BZ because they relax this constraint and consequently have to renormalize the resulting
matrices. This works well for $\mathrm{SL}(2,\reals)$ but already for $\mathrm{SL}(3,\reals)$ it gives spacetimes that do not
represent black holes. A clever solution of this problem in the BZ approach was found by Pomeransky \cite{Pomeransky}.  In this respect the BZ method supplemented with the Pomeransky trick  remains the most effective and powerful method for
constructing solutions of vacuum four- and five-dimensional gravity. For a
concise review and further references see \cite{Emparan:2008eg}. The Pomeransky trick works well for $\mathrm{SL}(n,\reals)$ but has no known
analog for other coset models. As mentioned in
the introduction, despite the initial complications, the promise of the BM method
lies in its generality; it can be taken over to other coset models.

It turns out that for $\mathrm{SL}(n,\reals)_\EE$ monodromy matrices with a maximum of two poles, it is rather straightforward to take
into account the coset constraints explicitly. When more poles are present one can
perhaps set up a recursive algorithm for finding the appropriate  meromorphic matrices.
We have not attempted this and leave this line of investigation for the future. Here we present a discussion of a two
soliton $\mathrm{SL}(2,\reals)_\EE$ matrix. The most general such configuration describes the Kerr-NUT solution as we show
now.

We start with the general form of $\cM_\EE(w)$ (compare (\ref{Mexp})),
\begin{align}
\label{bm2}
\cM_\EE(w) = \id + \frac{a_1 \alpha_1 a_1^T}{w - c} +  \frac{a_2 \alpha_2 a_2^T}{w + c},
\end{align}
where $a_1$ and $a_2$ are two-dimensional column vectors. The poles can be chosen in this way by a shift of axis, see (\ref{RSurface}).
Given the constant $2 \times 2$ matrix $a = (a_1, a_2)$ and the $2 \times 2$ matrix $\xi = a^T a$ we must choose
\begin{align}
\label{bm2const}
\alpha = \frac{2 c}{\det \xi} \left(
\begin{array}{cc}
\xi_{22}  & 0 \\
0         & - \xi_{11}
\end{array} \right),\quad\quad \alpha = \textrm{diag}\{\alpha_1,\alpha_2\},
\end{align}
in order to satisfy the constraint $\det \cM_\EE(w) = 1$. The matrix $\cM_\EE(w)$ is symmetric and is of determinant one, hence it is in the coset
$\mathrm{SL}(2,\reals)_\EE/\mathrm{SO}(2)_\EE$. For the parametrization of the inverse
$\cM_\EE(w)^{-1}$ we can choose $ b = a \xi^{-1} \epsilon $ and $\beta = -
\alpha \det \xi $ with
\begin{align}
\epsilon  =  \left(
\begin{array}{cc}
0  & -1 \\
1  & 0
\end{array} \right).
\end{align}
Due to the scaling freedom for the vectors, we can choose without any loss of generality
\begin{align}
a_1 = \left( \begin{array}{c}
1 \\
\zeta_1
\end{array} \right), \qquad \qquad
a_2 = \left( \begin{array}{c}
\zeta_2 \\
1
\end{array} \right).
\end{align}
Thus
\begin{align}
\xi = \left(
\begin{array}{cc}
1 + \zeta_1^2  & \zeta_1 + \zeta_2 \\
\zeta_1 + \zeta_2  & 1 + \zeta_2^2
\end{array} \right),
\end{align}
and
\begin{align}
\alpha_1 = \frac{2 c (1 + \zeta_2^2)}{(1-\zeta_1 \zeta_2)^2} \qquad \qquad \alpha_2 = -\frac{2 c (1 + \zeta_1^2)}{(1-\zeta_1 \zeta_2)^2}.
\end{align}
For the  $b_k$  vectors we have
\begin{align}
b_1 = \frac{1}{1-\zeta_1 \zeta_2}\left( \begin{array}{c}
-\zeta_1 \\
1
\end{array} \right), \qquad \qquad
b_2 =  \frac{1}{1-\zeta_1 \zeta_2} \left( \begin{array}{c}
-1 \\
\zeta_2
\end{array} \right),
\end{align}
and for $\beta_k$
\begin{align}
\beta_1  = - 2 c (1+ \zeta_2^2), \qquad \qquad \beta_2  =  2 c (1+ \zeta_1^2).
\end{align}
From the above expressions we see that $a_1^T  b_1 = 0 = a_2^T  b_2$,
as expected. Furthermore we have $a_2^T  b_1 = - a_1^T  b_2 = 1$. With the above choices we find
\begin{align}
\Gamma = \frac{1}{t_2 - t_1}
\left(
\begin{array}{cc}
\frac{\xi_{12}}{\xi_{22}} \frac{t_2 (t_1 + t_1^{-1})}{1 + t_1 t_2}  & 1 \\
1  & \frac{\xi_{12}}{\xi_{11}} \frac{t_1 (t_2 + t_2^{-1})}{1 + t_1 t_2}
\end{array} \right).
\end{align}
From $\Gamma$ we obtain the $c_k$ vectors by (\ref{solc}) and from there by looking at
the limiting value of the $A^{-1}_{+}(t)$ at $t = \infty$, cf.~equation \eqref{final2}, we obtain $M_{\mathrm{E}}(x)$.
We find
\begin{align}
M_\EE(x) = \id + a (\Gamma T \epsilon^{-1} \xi)^{-1} a^T, \qquad \mbox{where} \qquad T = \mbox{diag}\, \{t_1, t_2\}.
\end{align}
We also observe that
\begin{align}
\Gamma T \epsilon^{-1} \xi =(\alpha \nu)^{-1}
- \frac{t_1 t_2}{1 + t_1 t_2} \xi, \qquad \mbox{where} \qquad (\alpha \nu) = \mbox{diag}\, \{\alpha_1 \nu_1, \alpha_2 \nu_2\}.
\end{align}
From $M_{\mathrm{E}}(x)$ one can read off the physical fields. The conformal factor, given by (\ref{conformal_factor}), is
\begin{align}
\cf_\EE^2 =k_\mathrm{BM} \frac{t_1\nu_1 t_2\nu_2}{(t_2-t_1)^2} \left[ \frac{\xi_{12}^2}{\xi_{11}\xi_{22}} \frac{(1+t_1^2)(1+t_2^2)}{(1+t_1t_2)^2} -1\right].
\end{align}

\subsection{Interpretation as Kerr--NUT metric}

For four-dimensional vacuum gravity with $\mathrm{SL}(2,\reals)_\EE$ symmetry we can parametrise the monodromy $M_\EE$ as~\cite{Breitenlohner:1986um} %
\begin{align}
M_\EE = V_\EE^T V_\EE= \left(\begin{array}{cc}
\Delta+ \Delta^{-1}{\tilde\psi}^2& \Delta^{-1} \tilde\psi\\
\Delta^{-1}\tilde\psi & \Delta^{-1}
\end{array}\right).
\end{align}
Here, $\tilde\psi$ is dual to the metric function $\psi$ by the duality relation\footnote{From the point of view of the Geroch group, this duality with its non-linear prefactor is at the heart of the infinite-dimensional symmetry and integrability of the system: $\tilde\psi$ is the first in an infinite set of so-called dual potentials on which the infinite symmetry acts.}
\begin{align}
\label{dualityrelation}
\star_2 d\tilde \psi = - \frac{\Delta^2}\rho d\psi.
\end{align}
The $D=4$ metric is given by
\begin{align}
\label{4dmetric}
ds_4^2 =  - \Delta \left(dt+\psi d\phi \right)^2 + \Delta^{-1}(\cf_\EE^2\left(d\rho^2 + dz^2\right) + \rho^2 d\phi^2) .
\end{align}
(The form of the conformal factor is due to the change from Ehlers to Matzner--Misner variables to describe the physical space-time.)

To write explicit expressions for the scalars it is convenient to introduce prolate spheroidal
coordinates $(u,v)$
\begin{align}
z = u v, \qquad \rho = \sqrt{(u^2 -c^2)(1-v^2)}, \qquad c \le u < \infty, \qquad
-1 \le v \le 1.
\end{align}
These coordinates allow us to write the pole trajectories $t_1$ and $t_2$ as
\begin{align}
t_1 = \frac{(u-c)(1+v)}{\sqrt{(u^2 -c^2)(1-v^2)}}, \qquad \qquad t_2 =
\frac{(u+c)(1+v)}{\sqrt{(u^2 -c^2)(1-v^2)}}.
\end{align}
The inverse relations are
\begin{align}
\frac{t_1}{t_2} = \frac{u-c}{u+c},  \qquad \qquad t_1 t_2 = \frac{1 + v}{1-v}.
\end{align}
In these new coordinates we have
\begin{align}
\Delta =\frac{1}{D} \left[v^2 c^2 (\zeta_1 + \zeta_2)^2 + u^2 (1-\zeta_1 \zeta_2)^2 - c^2
(1+\zeta_1^2)(1 + \zeta_2^2)\right]
\label{Delta}
\end{align}
and
\begin{align}
\tilde \psi = \frac{1}{D}\left[
2 c u (\zeta_2 - \zeta_1) (1-\zeta_1 \zeta_2)
-2 c^2 v (\zeta_1 + \zeta_2) (1 + \zeta_1 \zeta_2)\right],
\label{tpsi}
\end{align}
where the common denominator of these expressions is
\begin{align}
D= v^2 c^2 (\zeta_1 + \zeta_2)^2
+ 2 v c^2 (\zeta_2^2 -\zeta_1^2)
+ u^2 (1 - \zeta_1 \zeta_2)^2 + c^2 (1 + \zeta_1^2) (1 + \zeta_2^2) + 2 c u (1 -
\zeta_1^2 \zeta_2^2).
\end{align}
By applying the duality relation (\ref{dualityrelation})
one can write an expression for $\psi$.\footnote{Alternatively, one could construct the generating function $\cV_\EE(t)$ and apply the algebraic Kramer--Neugebauer transformation
(relating the Ehlers and Matzner--Misner description) to obtain $\cV_\MM(t)$ that directly contains $\psi$~\cite{Breitenlohner:1986um}.}
It is slightly more
complicated
\begin{subequations}
\label{psi}
\begin{align}
\psi &= \frac{N_\psi}{(1 - \zeta_1 \zeta_2)D_\psi}, &\\
D_\psi &=  u^2 (1 - \zeta_1 \zeta_2)^2 -
   c^2 (1 + \zeta_1^2)(1 + \zeta_2^2) - c^2 v^2 (\zeta_1 + \zeta_2)^2, &\\
N_\psi
&=
- 4 c^3 \zeta_1 (1 + \zeta_1^2)(1 + \zeta_2^2)
- 2 c^2 (\zeta_1 + \zeta_2)(1- \zeta_1^2 \zeta_2^2) u
+ 2 c (\zeta_1 - \zeta_2)(1- \zeta_1 \zeta_2)^2 u^2 \nn \\
& \quad - 2 c^3 (\zeta_1 - \zeta_2)(1- \zeta_1 \zeta_2)^2 v
+ 2 c (\zeta_1 - \zeta_2)(1- \zeta_1 \zeta_2)^2 u^2 v \nn \\
& \quad + 2 c^3 (\zeta_1 + \zeta_2)(1 + 2 \zeta_1^2 + \zeta_1^2 \zeta_2^2) v^2
+ 2 c^2 (\zeta_1 + \zeta_2)(1 - \zeta_1^2 \zeta_2^2) u v^2.&
\end{align}
\end{subequations}
These expressions look somewhat cumbersome. To compare them with
the corresponding expressions for the Kerr--NUT metric in the standard
Boyer-Lindquist coordinates, let us recall that the latter takes the form (see e.g.~\cite{BV}) %
\begin{align}
ds^2 &=- \frac{1}{\Sigma} \left( \Xi - a^2 \sin^2 \theta \right) dt^2 +
\frac{\Sigma}{\Xi} dr^2 + \Sigma d\theta^2 + \frac{1}{\Sigma} \left( (\Sigma
+ a \chi)^2 \sin^2 \theta - \chi^2  \Xi \right) d\phi^2 \nn &\\
& \quad +
\frac{2}{\Sigma} \left( \chi \Xi - a (\Sigma + a \chi) \sin^2 \theta \right) dt d\phi,&
\label{KN_BL}
\end{align}
with
\begin{subequations}
\begin{align}
\Sigma &= r^2  + (n + a \cos \theta)^2  &\\
\Xi &= r^2 - 2 m r - n^2 + a^2 &\\
\chi &= a \sin^2 \theta - 2 n (1+ \cos \theta).&
\end{align}
\end{subequations}
The Boyer-Lindquist coordinates $(r,\theta)$ are related to the
prolate spheroidal coordinates (see e.g. appendix G of \cite{Harmark}) simply as,
\begin{align}
u = r -m, \qquad v = \cos \theta.
\end{align}
We find that with the identifications
\begin{align}
\zeta_1 = \frac{c-m}{a + n}, \qquad \qquad \zeta_2 = - \frac{a + n}{c + m},
\qquad \qquad c= \sqrt{m^2 + n^2 - a^2},
\label{zeta1zeta2}
\end{align}
the $tt$, $t \phi$, and $\phi \phi$ part of the metric
matches with corresponding expressions obtained through \eqref{Delta},
\eqref{tpsi} and \eqref{psi}.
Relations \eqref{zeta1zeta2} can be inverted to read
\begin{align}
a = -m\frac{\zeta_1+ \zeta_2}{1 + \zeta_1 \zeta_2}, \qquad \qquad n = m \frac{\zeta_1-\zeta_2}{1 + \zeta_1 \zeta_2}, \qquad m = c \frac{1 + \zeta_1 \zeta_2}{1 - \zeta_1 \zeta_2}.
\end{align}
When $\zeta_1 = \zeta_2$ we obtain the Kerr solution and the corresponding expressions match those of \cite{Maison} and \cite{BMunpub} (when certain minor typos and misprints are fixed in those references). When $\zeta_1 = \zeta_2=0$ we obtain the
Schwarzschild solution as in \cite{Maison, NicolaiLectures}.

The conformal factor can also be easily computed using the formula \eqref{conformal_factor}. We find
\begin{align}
\cf_\EE^2 = - k_\mathrm{BM} \frac{u^2 - m^2 -n^2 + a^2 v^2}{4 (m^2 + n^2) (u^2 -c^2 v^2)}.
\end{align}
Choosing the constant $k_\mathrm{BM}$ to be $-4 (m^2 + n^2)$ so that $f_{\mathrm{E}} \to 1$ as $r \to \infty$, we have
\begin{align}
\cf_\EE^2 = \frac{u^2 - m^2 -n^2 + a^2 v^2}{(u^2 -c^2 v^2)},
\end{align}
which in the MM coset allows us to match directly with the metric \eqref{KN_BL}
\begin{align}
\Delta^{-1} \cf_\EE^2 =  \frac{(m + u)^2 + (n + a v)^2}{u^2 -c^2 v^2}.
\end{align}

For completeness, we record the form of the (Ehlers) monodromy matrix $\cM_\EE(w)$ expressed in terms of the physical quantities
\begin{align}
\cM_\EE(w) = \frac{1}{w^2-c^2}
\left( \begin{array}{cc}
(m + w)^2 + (n+a)^2 & 2 (a m - n w)  \\
2 (a m - n w) &(w- m)^2 + (a - n)^2
\end{array} \right)
\end{align}
with $c$ as in (\ref{zeta1zeta2}).

\subsection{BZ Ehlers Construction}

The Kerr--NUT metric has been reconstructed by several authors in the context of
the standard BZ method, see e.g.~\cite{BV, Emparan:2008eg}. In this section we revisit this computation and perform it using the Ehlers generating function $\Psi_\EE(\lambda)$. To the best of our knowledge this has not been presented before.

The Killing part of the metric of flat space translates into $M_\EE$ being the
identity matrix. This implies that the seed $\Psi_{\EE}(\lambda, x)$ is also identity.
Now we add solitons at $\lambda = \mu_1$ with BZ vectors $m_0^{(1)}=(A_1, A_2)$ and at $\lambda = \mu_2$ with BZ vectors $m_0^{(2)} = (B_1, B_2)$. The $\Gamma_{\mathrm{BZ}}$ matrix is found to be
\begin{align}
\Gamma_{\mathrm{BZ}} = \left(
\begin{array}{cc}
 \frac{A_1^2+A_2^2}{\mu_1^2+\rho ^2} & \frac{A_1 B_1+A_2 B_2}{\rho ^2+\mu_1 \mu_2} \\
 \frac{A_1 B_1+A_2 B_2}{\rho ^2+\mu_1 \mu_2} & \frac{B_1^2+B_2^2}{\mu_2^2+\rho ^2}
\end{array}
\right)
\end{align}
and we need to rescale the resulting $M_\EE$ by $ - \frac{\mu_1 \mu_2}{\rho^2}$
\cite{BZ, BV} to ensure the determinant condition. With the choice of parameters
$w_1 = -c$ and $w_2= + c$ for the pole locations $\mu_1$ and $\mu_2$ in (\ref{BZpolepos}), and $(A_1, A_2) = (- \zeta_1, 1)$ and $(B_1, B_2) = (1, - \zeta_2)$ for the BZ vectors,
the final expression for the rescaled $M_\EE$ matches precisely with the
$M_\EE$ obtained from the BM method in section \ref{sec:KN}. The conformal factor obtained via equation \eqref{conformal_factorE} also matches with the one obtained
in section \ref{sec:KN}, with the constant $k_{\mathrm{BZ}}$ in \eqref{conformal_factorE} being $k_{\mathrm{BZ}} = 4 (m + c)^2$. It is intriguing to note that
 up to overall normalizations the BZ vectors are precisely the vectors $b_1$ and $b_2$
of the BM construction above.

\section{Conclusions}
\label{conclusions}

In this paper we studied the integrability of two-dimensional gravity-matter systems with matter related to a symmetric space $\mathrm{G}_\EE/\mathrm{K}_\EE$. The integrability
of these set-ups can be exhibited either using the Belinski--Zakharov (BZ) linear system or through the Breitenlohner--Maison (BM) linear system. As emphasized, the approach of
Breitenlohner--Maison makes the group structure of the Geroch group manifest. We analysed the relation between the BZ and the BM linear systems and presented explicit
relations between the generating functions appearing in these two linear systems.

An embedding of the Belinski--Zakharov solution generating
technique in the Geroch group was also studied in general. We pointed out that it is impractical to find a satisfactory general embedding of the full BZ
solution generating technique in the Geroch group. Relation \eqref{relationBZBM} provides in principle the link between the generating functions. However, one must keep in mind that the left hand side of \eqref{relationBZBM} must be
a `physical'  generating function $\Psi_\EE(\lambda)$. Here, by a `physical'  generating function we mean a generating function that gives a representative of the coset $\mathrm{G}_\EE/\mathrm{K}_\EE$ upon taking the limit $\lambda \to 0$.
This does not happen automatically in the BZ technique where the dressing matrix $\chi(\lambda)$ is considered in a more general context. %

On the other hand, following the unpublished work of Breitenlohner and Maison \cite{BMunpub}, we exhibited a novel solution generating method where the
group theoretical interpretation is clear from the beginning to the end. In our approach we solve the requisite Riemann--Hilbert problem algebraically. Since only algebraic manipulations are involved, our technique
 is akin to the BZ technique.  As a novel example, we constructed the Kerr-NUT solution in this approach. %

Our main interest in performing the analysis of integrability in these gravity-matter systems is to make the means  for constructing new solutions available in situations where the standard inverse scattering method of BZ is not applicable. This is typically the case for extended supergravity theories that have a string theory origin. For minimal $D=5$ supergravity with exceptional Ehlers symmetry $\mathrm{G}_\EE=\mathrm{G}_{2(2)}$ this problem was pointed out in~\cite{Figueras:2009mc} and also arises for the STU model \cite{Duff:1995sm} with $\mathrm{G}_\EE=\mathrm{SO}(4,4)$ or maximal supergravity with $\mathrm{G}_\EE=\mathrm{E}_{8(8)}$ \cite{Breitenlohner:1987dg}. The method explained in section~\ref{sec:solitons} is still applicable in those cases as long as one finds a way to parametrise the $\mathrm{G}_\EE$ valued monodromy $\cM_\EE(w)$ in a way similar to (\ref{bm2}) and (\ref{bm2const}). There are several ways in which this could be achieved: $(i)$ One could use global elements $k\in\mathrm{K}_\EE\subset \mathrm{G}_\EE\subset$ (Geroch group) to rotate the vectors in $\cM(w)$ in (\ref{bm2}) into canonical positions. In the example of section~\ref{sec:KN} this would correspond to setting $\zeta_1=\zeta_2$. The solution for canonical vectors can then be generalised by applying conventional $\mathrm{K}_\EE$ charging transformations. These charging transformations use only a very small subset of the full power of the Geroch group. $(ii)$ One could embed $\mathrm{G}_\EE$ in $\mathrm{GL}(n,\reals)$ for $n$ large enough and then solve the constraints on the vectors for the embedding explicitly. We expect a combination of these two techniques to be the most promising line of attack.
Alternatively to $(i)$ and $(ii)$, one could perhaps hope to develop some general algorithms to find the appropriate
monodromy matrices by combining ideas from  uniqueness proofs for black holes, see e.g.~\cite{Hollands:2012cc}, and the fact that the monodromy matrices are closely related
to the behavior of solutions on the z-axis, see e.g.~section 4 of \cite{Breitenlohner:1986um}. In future work we plan to explore these sets of ideas.
See also \cite{Shabnam} for a slightly different but related viewpoint on this problem.
It will be very interesting to see
if the algebraic Riemann--Hilbert factorization approach can be  used to construct
new black hole solutions generalizing \cite{Elvang:2004xi, Compere:2010fm} and new fuzzball solutions
generalizing~\cite{Jejjala:2005yu, Bena:2009qv}.

Another interesting aspect of the integrable structure in two-dimensional models is their possible relation to the recently studied infinite-dimensional symmetries of string and M-theory~\cite{West:2001as,Damour:2002cu}. These symmetry groups are extensions of the Geroch group and are conjectured to be symmetries of the unreduced theory. First steps in investigating this relation were undertaken in~\cite{Kleinschmidt:2005bq} in the restricted case of polarized Gowdy space-times. The relation (\ref{partialt}) between a spectral parameter and a space-time coordinate suggests that a mapping between Lie algebraic and geometric data might be possible. If taken seriously, this approach would allow the treatment of space-time as a concept that fully emerges from symmetry considerations.

\subsection*{Acknowledgements}
We thank Hermann Nicolai and especially Guillame Bossard for providing us with a
set of unpublished notes of Breitenlohner and Maison from June 1986. We are grateful for discussions with Shabnam Beheshti, Chand Devchand and Hermann Nicolai.

\appendix

\section{Computation of the BM conformal factor}
\label{conformal_factor:APP}

In this appendix we present a derivation of equation  \eqref{conformal_factor} for meromorphic monodromy matrices
following \cite{BMunpub}. Using the light cone coordinates \eqref{lightcone}
with the property $\star_2 \partial_{\pm} = \pm i \partial_{\pm}$ we can write the differential equations (\ref{confconstraint}) for the conformal factor $\cf_\EE$ as
\begin{align}
\partial_{\pm} \ln \rho \,  \partial_{\pm} \ln \cf_\EE = \frac{1}{2} \mbox{Tr} (P_{\EE, \pm} P_{\EE, \pm}).
\label{conf_eq}
\end{align}
Here, we have written the invariant bilinear form $\langle\cdot|\cdot\rangle$ as a matrix trace with a view to the application of the formalism to $SL(2,\reals)$.
Next we wish to write $ \mbox{Tr} (P_{\EE, \pm} P_{\EE, \pm})$ in
terms of the matrix $A_{+}(t)$ introduced in (\ref{finalcVE}) in section \ref{sec:solitons}. To this end we
evaluate the residue of the poles at $t = \pm i$ in the Lax equation \eqref{LSgen}. For evaluating the residue on the l.h.s. of (\ref{LSgen}) we use the relation
\begin{align}
\partial_{\pm} \cV_{\EE}(t,x)  = \partial_{\pm} \cV_{\EE}(t,x) |_{t} + (\partial_{\pm} t) \dot \cV_{\EE}(t,x),
\end{align}
where $\dot \cV_{\EE}(t,x) = \frac{\partial \cV_{\EE}(t,x)}{\partial t}$. These relations together with \eqref{partialt} give
\begin{align}
\pm i \partial_{\pm} \ln \rho \, \dot \cV_{\EE}(\pm i) = P_{\EE, \pm} \cV_{\EE}(\pm i). \label{elim}
\end{align}
Now replacing \eqref{finalcVE}  in \eqref{elim} we obtain an expression for $P_{\EE, \pm}$ in terms of $A_{+}(\pm i)$ and $\dot A_{+}(\pm i)$. Substituting
that expression in \eqref{conf_eq} we obtain
\begin{align}
\partial_{\pm} \ln \cf_{\EE} =  - \frac{1}{2}
(\partial_{\pm} \ln \rho) \mbox{Tr}\left(A_{+}^{-1}(\pm i) \dot A_{+}(\pm i)\right)^2.
\label{conf_eq2}
\end{align}
Using the explicit form of $A_{+}(t)$ from equation \eqref{Vplus} together with \eqref{solc} and the identity $a^T_k
b_k =  \Gamma_{kl} (t_k - t_l)$ we get,
\begin{align}
\label{dAplus}
A_{+}^{-1}(t) \dot A_{+} (t) = -  b \frac{\id}{\id + t T}
\Gamma^{-1}\frac{\id}{\id + t T} a^{T}.
\end{align}
In writing this equation we have used a convenient matrix notation, where $T$ is a diagonal matrix with entries $t_k$. Differentiating for $k\neq l$ the identity (cf. (\ref{Gammadef}))
\begin{align}
a^T_k
b_l =  \Gamma_{kl} (t_k - t_l),
\label{identity}
\end{align}
with respect to the light cone coordinates we obtain the components with $k \neq l$ of the equation
\begin{align}
\label{dgamma}
\partial_{\pm}\Gamma = -(\partial_{\pm} \ln \rho) \frac{\id}{\id \pm i T} \left[ \Gamma \mp i T \Gamma \mp i \Gamma T + T \Gamma T\right]\frac{\id}{\id \pm i T}.
\end{align}
Looking at the definition of the diagonal components $\Gamma_{kk}$
in (\ref{Gammadef}), we note that we need $\partial_\pm \gamma_k$ in order to obtain the corresponding expression for the diagonal components of $\Gamma$. We hence differentiate
the relation $\cA_k b_k = \gamma_k \alpha_k\nu_k a_k$ from \eqref{nosinglepole2}. An important intermediate result for this is that $\cA_k b_k$ is constant. From this it is easy to deduce that (\ref{dgamma}) holds as well for the diagonal components.

Using these formulas, we first substitute (\ref{dAplus}) into (\ref{conf_eq2}) and then manipulate the new r.h.s. to bring out terms that are total derivatives using (\ref{dgamma}). As a result, we can rewrite equation \eqref{conf_eq2} in the form
\begin{align}
\partial_{\pm} \ln \cf_\EE = \frac{1}{2} \mbox{Tr}\left(\Gamma^{-1}\partial_{\pm} \Gamma\right) + \frac{1}{2}\mbox{Tr}\left((T \nu)^{-1}\partial_{\pm} (T \nu)\right),
\label{conf_eq3}
\end{align}
where $(T \nu)$ is the diagonal matrix with entries $t_k \nu_k$. Equation \eqref{conf_eq3} can now be readily integrated to give the final result \eqref{conformal_factor}
\begin{align}
\cf_\EE^2 = k_\mathrm{BM} \cdot \prod_{k=1}^{N} (t_k \nu_k) \cdot \det \Gamma,
\end{align}
with $k_\mathrm{BM}$ an integration constant. (More generally, the conformal factor is related to a cocycle calculation in the affine group~\cite{Breitenlohner:1986um}.)

\section{BZ Matzner--Misner}
\label{app:BZMM}

The Belinski--Zakharov (BZ) approach is a well established solution generating
technique for vacuum gravity. The method is applicable in any dimension, though
only in four and five dimensions can the generated solutions be asymptotically
flat \cite{Emparan:2008eg}. In this appendix we focus on $D=4$. We assume that the
space-time admits two commuting Killing vectors,  one spacelike (angular) and one timelike. In this case the four-dimensional metric admits the following form in
the Weyl canonical coordinates (cf.~(\ref{4dmetric}))
\begin{align}
ds^2 = e^{2 \nu} \left(d\rho^2 + dz^2\right)+g_{ab} dx^a dx^b ,
\label{metric4dNEW}
\end{align}
where the indices $a, b$ run over the Killing coordinates $\phi$ and $t$. The vacuum Einstein equations can be used to choose without any loss of generality the coordinates such that \cite{Harmark}
\begin{align}
\det g = - \rho^2.
\end{align}

For this class of metrics, the Einstein equations divide in two groups: one for the Killing part $g$ of the metric
\begin{align}
\partial_\rho U + \partial_z V = 0, \qquad \mbox{with} \qquad U = \rho (\partial_\rho g) g^{-1} \quad \mbox{and} \quad V = \rho (\partial_z g) g^{-1}
\label{eqUV}
\end{align}
and the second group for the conformal factor $\nu$. From the discussion in the main text of the paper it
is clear that equation \eqref{eqUV} are the equations for a GL(2,$\reals$)
integrable sigma model. The equations for the conformal factor $\nu$ read
\begin{align}
\partial_\rho \nu = - \frac{1}{2 \rho} + \frac{1}{8 \rho} \mbox{Tr}(U^2-V^2), \qquad \qquad \partial_z \nu = \frac{1}{4 \rho} \mbox{Tr}(UV).
\label{eqnsconf}
\end{align}
Note that the equations for $g$ do not contain the function $\nu$. Once one obtains a solution of (\ref{eqUV}), $\nu$ can be obtained by a line integral.

The Belinski--Zakharov spectral equations for the above GL(2,$\reals$) model \eqref{eqUV} are~\cite{BZ}\footnote{Note that there is neither an Ehlers nor a Matzner--Misner subscript on this generating function $\Psi$ since it agrees with neither. It is, however, closely related to the Matzner--Misner version as will become clear in the sequel.}
\begin{align}
\label{BZlax}
D_1 \Psi = \frac{\rho V - \lambda U}{\lambda^2 + \rho^2} \Psi, \qquad \qquad D_2 \Psi = \frac{\rho U + \lambda V}{\lambda^2 + \rho^2}\Psi,
\end{align}
where $\lambda$ is the spectral parameter and $D_1$ and $D_2$ are the two commuting differential operators of (\ref{BZD1D2}).\footnote{In `light-cone' coordinates the spectral equations read
\begin{align}
D_{\pm} \Psi =\frac{\pm i\rho\left(\partial_{\pm}g\right)g^{-1}}{\lambda\pm i\rho}\Psi, \qquad \text{with} \qquad D_{\pm}=\partial_{\pm}-\frac{2\lambda}{\lambda\pm i\rho} \partial_\lambda.\nn
\end{align}}
The generating matrix $\Psi(\lambda, \rho, z)$ is such that in the limit $\lambda \to 0$ it gives the volumeful metric $g$. Using the above linear system \eqref{BZlax} one can construct an infinite class of new solutions by dressing seed solutions. This procedure has been reviewed at several places, see e.g.~\cite{BV,Emparan:2008eg}.

The matrix $g$ can also be written in terms of the $\mathrm{SL}(2,\reals)_\MM$ unimodular matrix $M_\MM$
\begin{align}
\label{volumeful}
 g=\rho M_\MM,
\end{align}
where the subscripts MM stand for Matzner--Misner. In terms of the coset variables it takes the form $M_\MM=V^T_\MM\eta V_\MM$, where $V_\MM$ is the Matzner--Misner coset representative of the quotient
$\mathrm{SL}(2,\mathbb{R})_\MM/\mathrm{SO}(1,1)_\MM$ and $\eta$ is the $\mathrm{SO}(1,1)_\MM$ invariant metric $\eta = \mbox{diag}\{1, -1\}$. Following \cite{Breitenlohner:1986um} we take the parameterization for $M_\mathrm{MM}$ to be
\begin{align}
M_\MM=V_\MM^T \eta V_\MM =\left(\begin{array}{cc}
\frac{\rho}{\Delta}-\frac{\Delta}{\rho}\psi^2 &-\frac{\Delta}{\rho} \psi\\
-\frac{\Delta}{\rho} \psi &- \frac{\Delta}{\rho}
\end{array}\right)
\quad\quad\text{with}\quad \quad
\eta =\left(\begin{array}{cc}
1&0\\0&-1
\end{array}\right),
\end{align}
where we emphasize that in our conventions the second index denotes the time component. We use this (non-standard) convention to facilitate comparison with \cite{Breitenlohner:1986um}.

Equations (\ref{eqUV}) and (\ref{eqnsconf}) can now be brought to the form
\begin{align}
\label{eqnsM}
 \partial_m\left(\rho M_\MM^{-1}\partial^m M_\MM\right)=0,
\end{align}
and
\begin{align}
\label{confMrho}
 \xi^{-1}\partial_\rho \xi=\frac{\rho}{8}\left(\text{Tr}\left(M_\MM^{-1}\partial_\rho M_\MM\right)^2-\text{Tr}\left(M_\MM^{-1}\partial_z M_\MM\right)^2\right),
\end{align}
\begin{align}
\label{confMz}
 \xi^{-1}\partial_z \xi=\frac{\rho}{4}\text{Tr}\left(M_\MM^{-1}\partial_\rho M_\MM M_\MM^{-1}\partial_z M_\MM\right),
\end{align}
where we have used $\square\rho=0$ and $\xi(\rho,z)$ is defined as $\xi=e^\nu \rho^{\frac{1}{4}}$. Defining as before
\begin{align}
P_{\MM,m} = \frac12 \left(\partial_m V_\MM V_\MM^{-1} + \left(\partial_m V_\MM V_\MM^{-1}\right)^T\right),
\end{align}
we can rewrite the above equations in the coordinates $x^{\pm}$ of (\ref{lightcone}) as
\begin{subequations}
\label{2deomsBZ}
\begin{align}
\label{confxi}
\pm i \xi^{-1}  \partial_\pm\xi &= \frac{\rho}{2} \text{Tr}\left( P_{\mathrm{MM},\pm} P_{\mathrm{MM},\pm}\right),\\
\label{2dmatterBZ}
D_m(\rho\, P_{\mathrm{MM}}^m) &=0.
\end{align}
\end{subequations}
Equations (\ref{confxi}) and (\ref{2dmatterBZ}) are formally identical to equations
\eqref{confconstraint} and \eqref{2dmatter}.
The BZ Lax pair can be written for these equations \cite{Figueras:2009mc} as well.

If one wants to relate the BZ-generating function $\Psi(\lambda)$ to the group-theoretic BM-generating function $\cV_\MM(t)$ additional care has to be taken because of the factor of $\rho$ in (\ref{volumeful}). A convenient choice is
\begin{align}
\Psi(\lambda,x)  = \sqrt{2\rho t w} V^T_\MM(x) \eta \cV_\MM(t,x).\label{relationBZBMMM}
\end{align}
(Note that this differs from what was given in~\cite{Breitenlohner:1986um}; their choice does not map the linear systems into each other away from $t=\lambda=0$.)

\end{document}